\documentclass[12pt]{article}
\usepackage{epsfig}
\usepackage{tabularx}
\usepackage[titletoc, title]{appendix}
\usepackage{setspace}
\usepackage{lscape}
\usepackage{booktabs}
\usepackage{array}
\usepackage{makeidx}
\usepackage{latexsym}
\usepackage{amssymb}
\usepackage{qsymbols}
\usepackage{graphicx}
\usepackage{arydshln}
\usepackage{xr}
\usepackage{amsmath, amsthm, amssymb}
\usepackage{nomencl}
\usepackage{float}
\usepackage{multirow}
\usepackage{multicol}
\usepackage{xcolor,soul}
\usepackage{lscape}
\usepackage{tabularx}
\usepackage{pdflscape}
\usepackage{epstopdf}
\usepackage{color}
\usepackage{dsfont}
\usepackage[authoryear]{natbib}
\usepackage{mathrsfs}
\usepackage{listings}
\usepackage{spverbatim}

\lstnewenvironment{rc}[1][]{\lstset{language=R}}{}

\let\proglang=\textsf

\def\b#1{\mbox{\boldmath $#1$}}    

\title{\texttt{lqmix}: an \textsf{R} package for longitudinal data analysis via linear quantile mixtures}

\author{Marco Alf\`o\thanks{Dipartimento di Scienze Statistiche, Sapienza Universit\`a di Roma} \and Maria Francesca Marino\thanks{Dipartimento di Statistica, Informatica, Applicazioni, Universit\`a  degli Studi di Firenze. \texttt{mariafrancesca.marino@unifi.it}} \and Maria Giovanna Ranalli\thanks{Dipartimento di Scienze Politiche, Universit\`a  degli Studi di Perugia} \and Nicola Salvati\thanks{Dipartimento di Economia e Management, Universit\`a di Pisa}}

\date{}

\begin{document}



\maketitle

\abstract{
The analysis of longitudinal data gives the chance to observe how unit behaviors change over time, but it also poses a series of issues. These have been the focus of an extensive literature in the context of linear and generalized linear regression moving also, in the last ten years or so, to the context of linear quantile regression for continuous responses. In this paper, we present \texttt{lqmix}, a novel \texttt{R} package that  assists in estimating a class of linear quantile regression models for longitudinal data, in the presence of time-constant and/or time-varying, unit-specific, random coefficients, with unspecified distribution. Model parameters are estimated in a maximum likelihood framework via an extended EM algorithm, while parameters' standard errors are derived via a block-bootstrap procedure. The analysis of a benchmark dataset is used to give details on the package functions.
}
\section{Introduction}\label{sec:intro}
Quantile regression \citep{KoenkerBassett1978} has become quite a popular technique to model the effect of observed covariates on the conditional quantiles of a continuous response of interest. 
This represents a well established practice for the analysis of data when the focus goes beyond the conditional mean. Comprehensive reviews on this topic can be found, among others, in \cite{koen:hall:01}, \cite{yu:lu:stan:03}, \cite{Koenker2005}, and \cite{Hao2007}. 

With longitudinal data, dependence between observations recorded from the same statistical unit needs to be taken into account to avoid bias and inefficiency in parameter estimates. Different modeling alternatives are available in the literature for handling such a dependence. For a general presentation, see e.g., \cite{Diggle2002} and \cite{Fitzmaurice2012}, while for a focus on quantile regression see \cite{MarinoFarcomeni2015}. Here, we consider quantile regression models that include unit-specific random coefficients to describe such a dependence; this gives rise to a conditional model specification which allows one to draw unit-level inferential conclusions on the effect of covariates on the longitudinal outcome of interest. 
In this framework, a standard way of proceeding is based on specifying a parametric distribution for the random coefficients, as suggested, e.g., by \cite{GeraciBottai2007, GeraciBottai2014}. An alternative consists in leaving the distribution unspecified and estimating it from the observed data, by using a finite mixture specification. This can arise from discrete, unit-specific, random coefficients with unspecified distribution that remains constant or evolves over time according to a hidden Markov chain, as proposed by \cite{AlfoRanalliSalvati2016} and \cite{Farcomeni2010}, respectively. A more flexible specification based on combining both time-constant (TC) and time-varying (TV), unit-specific, discrete, random coefficients may also be adopted, as proposed by  \cite{Marino2018}.
When compared to fully parametric alternatives, this semi-parametric approach offers a number of specific advantages, as it helps (\textit{i}) avoid unverifiable assumptions on the random coefficient distribution; (\textit{ii}) account for extreme and/or asymmetric departures from the homogeneous model; (\textit{iii}) avoid integral approximations and, thus, considerably reducing the computational effort for parameter estimation.

In this paper, we describe the \textsf{R} package \citep{Rteam2019}  {\texttt{lqmix}} \citep{lqmixpack}, available from the Comprehensive \textsf{R} Archive Network (CRAN) at \texttt{https://\\CRAN.R-project.org/package=lqmix}, which is intended to provide maximum likelihood (ML) estimates for TC and/or TV mixtures of linear quantile regression models for longitudinal data. An indirect estimation approach, based on an extended Expectation-Maximization algorithm \citep[EM -][]{Dempster1977} is employed.

The package \texttt{lqmix} features with some relevant \textsf{R} packages available for the analysis of longitudinal data on the CRAN repository. It is related to packages \texttt{lme4} \citep{lme4Pack} and \texttt{lqmm}  \citep{lqmmPack} tailored, respectively, to the analysis of general clustered observations via mixed models for the conditional mean and the conditional quantiles of a response. Here, TC random coefficients following a parametric distribution are considered to model dependence between observations and a ML approach is employed to derive parameter estimates. %
\texttt{lqmix} is also strongly related to the \texttt{LMest} \textsf{R} package \citep{LMestPack}. This allows to model the mean of longitudinal observations via hidden Markov models \citep{Zucchini2009, BartFarcoPennoni2013}. In this case, TV random coefficients with unspecific distribution that evolves over time according a Markov chain are considered to account for dependence between measures from the same unit and a  ML approach is adopted to derive parameter estimates. 
Other \textsf{R} packages that can be fruitfully related to \texttt{lqmix} are \texttt{rqpd} \citep{rqpdPack}, \texttt{pqrfe} \citep{pqrfePack}, and \texttt{npmlreg} \citep{npmlregPack}. The former allows for the estimation of linear quantile regression models for panel data by considering either a penalized fixed effect estimation \citep{Koenker2004} or a correlated random-effect method \citep{AbrevayaDahl2008}. In both cases, parameters are estimated by minimizing an extended quantile loss function. Similarly, \texttt{pqrfe} allows for the estimation of quantile regression models for longitudinal data based on unit-specific fixed effects; three different estimation methods are implemented, each based on the minimization of a different loss function. Last, the \texttt{npmlreg} \textsf{R} package entails mixtures of (generalized) linear models for clustered observations by employing a ML approach for parameter estimation. Finally, it is worth mentioning  the \texttt{quantreg} \textsf{R} package \citep{quantregPack} for estimation and inference in linear models for conditional quantiles. Comparing these alternative packages with \texttt{lqmix}, it is worth highlighting that the latter fills in a blank by providing modeling tools not available in these other packages and ensures greater flexibility thanks to the non-parametric nature of the random coefficient distribution, which allows to avoid untestable parametric assumptions. The second benefit one could have from \texttt{lqmix} entails the modeling of quantiles rather than the mean. Quantile regression allows indeed to analyze the impact that predictors may have on different parts of the conditional response distribution, as well as, to deal with outliers and/or heavy tails that make the Gaussian assumption typically used for continuous data unreliable.

The modeling of longitudinal data via quantile regression is also possible by considering packages and environments out of the \textsf{R} world, although none of them allows to consider random coefficients and the possibility of modeling them {non-parametrically and/or} dynamically. The  \textsf{Stata} modules \texttt{xtqreg} \citep{xtqreg_pack} and \texttt{xtmdqr} \citep{xtmdqr_pack} allow for the estimation of linear quantile regression models for longitudinal data based on the use of fixed effects. These modules differ in the way model parameters are estimated: the former considers the method of moments introduced by  \cite{MachadoSantosSilva2019}, while the latter builds up on the  minimum distance approach described by \cite{GalvaoWang2015}. The \texttt{qregpd} module \citep{qregpd_pack} is also worth to be mentioned; it implements the quantile regression estimator developed in \cite{Graham2015} for panel data. Last, the \texttt{qreg} \textsf{Stata} module \citep{qreg2_pack} allows for the estimation of linear quantile regressions under the assumption of independent observations. Similarly does also the \texttt{quantreg} \textsf{SAS} procedure. 

The paper is structured as follows. 
The different proposals available in the literature on finite mixtures of linear quantile regression models are described in Section ``\textit{Modeling alternatives}'', 
 where the case of TC, TV, and the combination of TC and TV random coefficients are discussed. ML estimation is reviewed in Section  ``\textit{Model estimation and  inference}'',
 where details of the EM algorithm and the bootstrap procedure for standard error estimation are described. The proposed \textsf{R} package is presented in Section ``\textit{The \texttt{lqmix}  \textsf{R} package}'',
 where the analysis of a benchmark dataset is discussed. The last section contains some concluding remarks.

\section{Modeling alternatives}\label{sec:models}

{Consider the case in which a} longitudinal study is planned to record the value of a continuous response variable $Y$ and a set of 
covariates (or predictors or explanatory variables)
on a sample of $n$ statistical units, at $T$ different measurement occasions. This is a balanced study, with common (at least in unit-time) and equally spaced occasions, in discrete time. Let $Y_{it}$ denote the value of the response for unit $i=1, \dots, n,$ at occasion $t = 1, \dots, T,$ and $y_{it}$ the corresponding realization
for $i = 1, \dots, n,$ and $t = 1, \dots, T$ . Further, let $\b Y_i = (Y_{i1}, \dots, Y_{iT})^\prime$ and $\b y_i = (y_{i1}, \dots, y_{iT})^\prime$ 
be the $T$-dimensional vector of responses and its realization, respectively.
A frequent issue to address when dealing with longitudinal studies is that of missingness. This may entail both the response and the covariates. Here, we assume that the latter are fully observed, while an ignorable missingness \citep{Rubin1976} may affect the outcome. That is, some units in the sample may present incomplete response sequences due to either monotone or non-monotone missing data patterns \citep{LittleRubin2002}. In this sense, a varying number of measures $T_i$ may be available for each unit, even though missingness is assumed to be independent from unobserved responses -- M(C)AR assumption.

Random coefficient models represent a standard approach to analyze the effect of observed covariates on a response $Y$ that is repeatedly observed over time. This also holds in the quantile regression framework, where the interest is on modeling the conditional quantiles of the response distribution as a function of fixed and random coefficients. To ensure flexibility and avoid unverifiable parametric assumptions on the random coefficient distribution, a specification based on finite mixtures represent a viable strategy to adopt. For this purpose, we developed the \texttt{lqmix}  \textsf{R} package.

In this section, we describe the methodology underlying the proposed package, and present some alternative formulations of quantile regression models available in the literature to deal with longitudinal data. In detail, we consider models based on discrete, unit-specific, random coefficients with unspecific distribution able to capture unit-specific sources of unobserved heterogeneity due to omitted covariates. According to the chosen specification, these may remain constant and/or evolve over time, leading to a model based on TC, TV, or both TC and TV random coefficients, respectively. 
The following sections present each of these formulations in detail. 
%
%

\subsection{Linear quantile mixtures with TC random coefficients}\label{sec:modelsTC}
For a given quantile level $q \in(0,1)$, let $\b\beta_q$ denote a quantile-dependent, $p$-dimensional, vector of parameters associated with the (design) vector $\b x_{it}= (x_{it1}, \dots, x_{itp})^{\prime}$. Also, let $\b z_{it} = (z_{it1}, \dots, z_{itd})^\prime$ denote a set of $d \ge 1$ covariates (not included in $\b x_{it}$) associated with a vector of unit- and quantile-specific random coefficients $\b b_{i,q}=(b_{i1,q}, \dots, b_{id,q})^\prime$. 
The latter account for unobserved heterogeneity that is not captured by the elements in $\b x_{it}$ and may be used to describe dependence between repeated measurements from the same unit. In this sense, conditional on $\b b_{i,q}$, the longitudinal responses from the same unit, $Y_{i1}, \dots, Y_{it}$, are assumed to be independent ({local independence assumption}) {of} each other. In detail, for a given quantile level $q \in (0,1)$ and conditional on the vector $\b b_{i,q}$, the response $Y_{it}$ is assumed to follow an Asymmetric Laplace Distribution \citep[ALD - e.g.,][]{YuMoyeed2001}, with density
\[
f_{y \mid b} (y_{it} \mid \b b_{i,q}; q) = 
\left[\frac{q (1-q)}{ \sigma_q}\right] \exp \left\{ -\rho_q \left[ \frac{ y_{it} - \mu_{it,q}
}
{\sigma_q} \right] \right\}.
\]
Here, $\rho_q(\cdot)$ denotes the quantile asymmetric loss function \citep{KoenkerBassett1978}, while $q, \sigma_q$, and $\mu_{it,q}$ denote the skewness, the scale, and the location parameter of the distribution, respectively. The ALD is a working model that is used to recast estimation of parameters for the linear quantile regression model in a maximum likelihood framework. The location parameter of the ALD, $\mu_{it,q}$, is modeled as
\begin{equation}\label{eq:TCmodel_Par}
\mu_{it,q} = \b x_{it}^\prime \b \beta_q + \b z_{it}^\prime \b b_{i,q}.
\end{equation}

The modeling structure is completed by the mixing distribution $f_{b,q}({\b b}_{i,q}; {\b \Sigma}_{q})$, i.e., the distribution of the random coefficients ${\b b}_{i,q}$, where $\b \Sigma_q$ identifies a (possibly) quantile-dependent covariance matrix. 
Rather than specifying such a distribution parametrically as in, e.g., \cite{GeraciBottai2014}, \cite{AlfoRanalliSalvati2016} proposed to leave it unspecified and use a NonParametric Maximum Likelihood approach \citep[NPML --][]{Laird1978, Lindsay1983a, Lindsay1983b} to estimate it directly from the observed data. This approach is known to lead to the estimation of a (quantile-specific) discrete mixing distribution defined over the set of locations $\{\b \zeta_{1,q}, \dots, \b \zeta_{G_q,q}\}$, with mixture probabilities $\pi_{g,q} = \Pr(\b b_{i,q} = \b \zeta_{g,q})$, $i = 1, \dots, n, \, g = 1, \dots, G_q$, and $G_q\leq n$. 
Under this approach, for $\b b_{i,q}= \b \zeta_{g,q}$, the location parameter of the ALD in Equation \eqref{eq:TCmodel_Par} becomes
\begin{equation*}
\mu_{itg,q} = \b x_{it}^\prime \b \beta_q + \b z_{it}^\prime \b \zeta_{g,q},
\end{equation*}
while the model likelihood is defined by the following expression:
\begin{equation}\label{eq:TClk}
L(\cdot \mid q) =\prod_{i = 1}^{n} \sum_{g = 1}^{G_q} \left[\prod_{t = 1}^{T_i} f_{y\mid b}(y_{it} \mid \b b_{i,q} = \b \zeta_{g,q}; q)\right] \pi_{g,q}. 
\end{equation}
This equation clearly resembles the likelihood of a finite mixture of linear quantile regression models with TC, discrete, random coefficients. 

\subsection{Linear quantile mixtures with TV random coefficients}\label{sec:modelsTV}
While the model specification introduced in the previous {section} accounts for unit-specific omitted covariates,  it may fall short in handling time variations {in such unobserved heterogeneity. To address this limitation} \cite{Farcomeni2010} introduced a linear quantile regression model with TV, discrete, random intercepts. Rather than allowing the distribution of these intercepts to vary freely, it is modeled via a discrete-time Markov chain to ensure parsimony and interpretability.
While such a proposal entails unit-specific intercepts only, a broader specification with general TV, discrete, random coefficients can also be considered.
 
As before, let $\b x_{it}= (x_{it1}, \dots, x_{itp})^{\prime}$ denote a vector of  $p$ covariates associated with the vector of parameters $\b \beta_{q} = (\beta_{1,q}, \dots, \beta_{p,q})^{\prime}$. Further, let $\b w_{it} = (w_{it1}, \dots, w_{itl})^\prime$ denote a set of $l\geq 1$ explanatory variables not included in $\b x_{it}$, associated with a vector of unit-, time-, and quantile-specific random coefficients $\b \alpha_{it,q} = (\alpha_{it1,q}, \dots, \alpha_{itl,q})^\prime$. These are assumed to evolve over time according to  a homogeneous, first order, hidden Markov chain $\{S_{it,q}\}$ that depends on the specified quantile level $q$.  In detail, $\{S_{it,q}\}$ is defined over the finite state space $\mathscr S_q = \{1, \dots, m_q\}$, with initial and transition probabilities $\delta_{s_{i1},q}$ and $\gamma_{s_{it}|s_{it-1},q}$, defined as:
\[
\delta_{s_{i1},q} = \Pr(S_{i1,q} = s_{i1,q})
\]
and
\[
\gamma_{s_{it,q}|s_{it-1,q},q} = \Pr(S_{it,q} = s_{it,q}\mid S_{it-1,q} = s_{it-1,q}).
\] 

As for the  model based on TC random coefficients detailed above, local independence holds, together with the convenient assumption of conditional ALD for the responses.
For a given quantile level $q \in (0,1)$, the joint conditional distribution of the observed $T_i$-dimensional response vector $\b y_i = (y_{i1}, \dots, y_{iT_i})^\prime$ is given by
\[
f_{y \mid s} (\b y_i \mid \b s_{i,q}; q) = f_{y \mid s}(y_{i1} \mid  {s_{i1,q}}; q) \prod_{t = 2}^{T_i} f_{y \mid s}(y_{it} \mid  s_{it,q}; q).
\]
Here, $\b s_{i,q} = (s_{i1,q}, \dots, s_{iT_i,q})^\prime$ is the vector of states visited by the $i$-th unit and $f_{y \mid s}(y_{it} \mid  s_{it,q}; q)$ denotes the ALD with skewness $q$, scale $\sigma_q$, and location parameter $\mu_{its_{it,q}, q}$. This latter is modeled as
\[
\mu_{its_{it, q},q} = \b x_{it}^\prime \b \beta_q + \b w_{it}^\prime \b \alpha_{s_{it,q},q}.
\]
According to the assumptions  above, the likelihood function is
\begin{equation*}
L(\cdot \mid q) = \prod_{i = 1}^{n} \sum_{s_{i1}}^{m_q}\cdots \sum_{s_{iT_i}=1}^{m_q} \left[\delta_{s_{i1},q} \prod_{t = 2}^{T_i} \gamma_{s_{it,q}|s_{it-1,q}}\right] \left[\prod_{t = 1}^{T_i} f_{y \mid s}(y_{it} \mid  {s_{it,q}}; q)\right].
\end{equation*}
When looking at the above expression, we may recognize it is a dynamic extension of  Equation \eqref{eq:TClk}, as it represents the likelihood of a dynamic finite mixture of linear quantile regression models with TV, discrete, random coefficients. These coefficients are here assumed to vary as a function of the hidden states visited by the observed units over the observed time window. 

\subsection{Linear quantile mixtures with both TC and TV random coefficients}\label{sec:modelsTCTV}

\begin{figure}
\centering 
\includegraphics[width=0.53\textwidth]{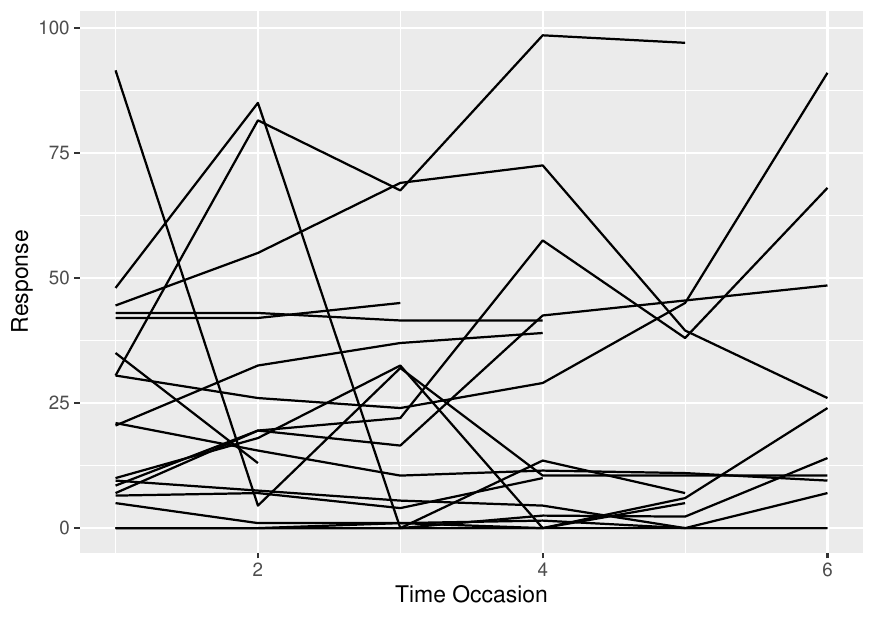}
\caption{An example of longitudinal trajectories affected by both time-constant and time-varying unobserved heterogeneity}\label{fig:trajectories}
\end{figure}

In some real data applications, both TC and TV sources of unit-specific unobserved heterogeneity may be present and influence the response distribution. We report in Figure \ref{fig:trajectories} the longitudinal trajectories representing the evolution of a given response measured over (at most) 6 time occasions for a sample of 25 statistical units. From this figure, it is clear that both sources of unobserved heterogeneity affect the response. TC unobserved features may be responsible of differences between units in terms of baseline levels and systematic temporal trends; TV unobserved features may instead explain the sudden temporal shocks characterizing individual trajectories. These latter may be rather difficult to capture via TC random coefficients or unit-specific random slopes associated to a time variable.
In these situations, the linear quantile mixtures described above are no longer appropriate, as they may account for one source at a time only. 
To model empirical cases where both sources of between and within-unit variation are present, \cite{Marino2018} introduced a linear quantile regression model where TC and TV random coefficients may be jointly present in the linear predictor.

Let $\b x_{it}= (x_{it1}, \dots, x_{itp})^{\prime}$ indicate a $p$-dimensional vector of covariates associated with the parameters $\b \beta_{q} = (\beta_{1,q}, \dots, \beta_{p,q})^{\prime}$. Furthermore, let $\b z_{it} = (z_{it1}, \dots, z_{itd})^\prime$ and $\b w_{it} = (w_{it1}, \dots, w_{itl})^\prime$ denote two disjoint vectors of $d \ge 1$ and $l\geq 1$ explanatory variables (not included in $\b x_{it}$), respectively. The former  is associated with the vector of unit- and quantile-specific random coefficients $\b b_{i,q}=(b_{i1,q}, \dots, b_{id,q})^\prime$ taking value in the in the set $\{\b \zeta_{1,q}, \dots, \b \zeta_{G_q,q}\}$ with probability $\pi_{g,q} = \Pr(\b b_{i,q} = \b \zeta_{g,q}), g = 1, \dots, G_q$. The latter is associated to the vector of unit-, time-, and quantile-specific random coefficients $\b \alpha_{it,q} = (\alpha_{it1,q}, \dots, \alpha_{itl,q})^\prime$, which evolves over time according to a homogeneous, first order, quantile-dependent, hidden Markov chain $\{S_{it, q}\}$. As before, this is defined over the finite state space $\mathscr S_q = \{1, \dots, m_q\}$ and is fully described by means of the initial probability vector $\b \delta_q =(\delta_{1,1}, \dots, \delta_{m_q,q})^\prime$ and the transition probability matrix $\b \Gamma_{q}$, with generic element $\gamma_{s_{it,q} \mid s_{it-1,q},q}.$
The reader must note that the intercept term is included in either $\b z_{it}$ or $\b w_{it}$ (or in neither), but never in both. A similar principle applies to unit-specific, discrete, random slopes in model, ensuring that $\b z_{it}$ and $\b w_{it}$ have no common elements.
Further, random coefficients $\b \zeta_{g,q}$ and $\b\alpha_{s_{it,q},q}$ are assumed to be independent.

For a given quantile level $q \in (0,1)$ and conditional on $\b b_{i,q} = \b \zeta_{g,q}$ and $\b \alpha_{it,q} = \b \alpha_{s_{it,q},q}$, longitudinal responses recorded on the same unit are assumed to be independent (local independence) and to follow an ALD with skewness, scale, and location parameter denoted by $q$, $\sigma_q$, and $\mu_{itgs_{it}, q}$, respectively. This latter parameter is defined by the following regression model: 
\[
\mu_{itgs_{it,q}, q} = \b x_{it}^\prime \b \beta_q + \b z_{it}^\prime \b \zeta_{g,q} + \b w_{it}^\prime \b \alpha_{s_{it,q},q}.
\]
Based on such assumptions, the likelihood function is 
\begin{align*} 
L(\cdot \mid q) =\prod_{i = 1}^{n}& \sum_{g = 1}^{G_q}  \sum_{s_{i1}}^{m_q}\cdots \sum_{s_{iT_i}=1}^{m_q} \left[
 \delta_{s_{i1,q}} \prod_{t = 2}^{T_i} 
\gamma_{s_{it,q} \mid s_{it-1,q}}\right] \left[\prod_{t = 1}^{T_i}  f_{y \mid b,s}(y_{it} \mid \b b_{i,q} = \b \zeta_{g,q}, {s_{it,q}};q)\right]
\pi_{g,q}, 
\end{align*}
where, as before, $f_{y \mid b,s}(y_{it} \mid \b b_{i,q} = \b \zeta_{g,q}, {s_{it,q}};q)$ denotes the density of the ALD. In this framework, unobserved unit-specific features that remain constant over time are captured by the random coefficients $\b \zeta_{g,q}$; sudden temporal shocks in the unit-specific profiles, due to TV sources of unobserved heterogeneity, are captured instead by the random coefficients $\b \alpha_{s_{it,q},q}$.

To conclude this section, it is worth noticing that, when a single hidden state ($m_{q}=1$) is considered and $\b b_{i,q}= \b \zeta_{g,q}$, the location parameter  $\mu_{itgs_{it,q},q}$ simplifies to $\mu_{itg,q}$ and the model reduces to the linear quantile mixture with TC random coefficients only. Also, when a single mixture component is considered ($G_{q} = 1$), the  location parameter $\mu_{itgs_{it,q},q}$ simplifies to $\mu_{its_{it,q},q}$ and the model above reduces to the dynamic finite mixture of linear quantile regressions with TV random coefficients only. Last, when both $G_{q}$ and $m_{q}$ are equal to 1, the model reduces to a standard linear quantile regression model, without random coefficients. These properties make this latter specification more flexible and general than the alternatives described so far, at the cost of a higher computational complexity.

\section{Model estimation and inference}\label{sec:ML}
In this section, we describe the algorithm for ML estimation of model parameters in  the linear quantile mixture models detailed in the previous section. We focus on the specification based on both TC and TV random coefficients, 
as the simpler alternatives based on TC or TV random coefficients only
can be derived as special cases.  We provide details of an EM algorithm for parameter estimation in Section ``\textit{Maximum likelihood estimation}'';
the procedure for deriving standard errors and choosing the optimal number of mixture components and/or states of the hidden Markov chain is described in Section ``\textit{Standard errors and model selection}''.

\subsection{Maximum likelihood estimation}\label{sec:MLest}
Let $\b \theta_{q}$ denote the global set of free model parameters for a given quantile level $q \in (0,1)$. To derive an estimate for such a vector, we may rely on indirect maximization of the likelihood function via an extended version of the EM algorithm \citep{Dempster1977}. Let $u_{i,q}(g)$ and $v_{it,q}(h)$ be the indicator variables for $\b b_{i,q} = \b \zeta_{g,q}$ and  $S_{it,q} = h$, respectively, and let $v_{it,q}(h,k) = v_{it-1,q}(h)\times v_{it,q}(k)$ denote the indicator variable for unit $i$ moving from state $h$ at occasion $t-1$ to state $k$ at occasion $t$, with $g = 1, \dots, G_{q},$ and $h, k = 1, \dots, m_{q}$. For a given $q \in (0,1)$, the EM algorithm starts from the following complete-data log-likelihood function: 
\begin{align}\label{eq:completeL}
\ell_c (\b \theta_{q} ) &= 
\sum_{i= 1}^n  \left\{
\left[ \sum_{g=1}^{G_q} u_{i,q}(g) \log \pi_{g,q}\right] \right.\\\nonumber 
&+
\left. \left[\sum_{h=1}^{m_q} v_{i1,q}(h)\log \delta_{h,q} +\sum_{t = 2}^{T_i} \sum_{h = 1}^{m_q} \sum_{k = 1}^{m_q} v_{it,q}(h,k) \log \gamma_{k\mid h,q} \right]  \right.\\\nonumber 
&+
\left.\left[\sum_{t = 1}^{T_i} \sum_{g =1}^{G_{q}}\sum_{h=1}^{m_{q}}   u_{i,q}(g) v_{it,q}(h) \log f_{y \mid b,s} (y_{it}\mid \b b_{i,q} = \b\zeta_{g,q}, S_{it,q} = h)\right] \right\}.
\end{align}
At the $r$-th iteration, the E-step of the EM algorithm requires the computation of the expected value of the complete-data log-likelihood in Equation \eqref{eq:completeL}, conditional on the observed data $\b y = (\b y_1, \dots, \b y_n)^\prime$ and the current parameter estimates $\b \theta_q^{(r-1)}$. That is, it requires the computation of 
$$Q(\b \theta_q \mid \b \theta_q^{(r-1)}) = \mbox{E}\left[\ell_c(\b \theta_{q}) \mid \b y, \b \theta_q^{(r-1)}\right].$$
 This means computing the posterior expectations of the indicator variables $u_{i,q}(g)$, $v_{it,q}(h)$, and $v_{it,q}(h,k)$. Regarding these two latter, a simplification is obtained by considering the forward and backward probabilities \citep{Baum1970, Welch2003} typically used in the hidden Markov model framework; see \cite{Marino2018} for further details.
In the M-step of the algorithm, parameter estimates $\hat{\b \theta}_q$ are derived by maximizing $Q(\b \theta_q \mid \b\theta_q^{(r-1)})$ with respect to $\boldsymbol \theta_q$.

The E- and the M-step are alternated until convergence, which is defined as the (relative) difference between two subsequent likelihood values being lower than a given threshold, $\varepsilon>0.$

\subsection{Standard errors and model selection}
\label{sec:SeModChoice}
Following the standard procedure used in the quantile regression framework, standard errors for model parameter estimates are derived by exploiting a nonparametric bootstrap approach \citep[see e.g.,][]{Buchinsky1995}. In detail, we employ a block-bootstrap procedure, where a re-sampling of the statistical units is performed and the corresponding sequence of observed measurements is retained to preserve within-unit dependence \citep{Lahiri1999}.

Let $\hat{\b \theta}^{(r)}_{q}$ denote the vector of model parameter estimates obtained in the $r$-th bootstrap sample, $r = 1, \dots, R$. Estimates of the standard errors for the vector $\hat{\b \theta}_{q}$ correspond to the diagonal elements of the matrix
\begin{equation*}
\hat{ \textbf V}(\hat{\b \theta }_{q}) =
 \sqrt{\frac{1}{R} \sum_{r = 1}^{R} \left( \hat{\b \theta}^{(r)}_{q} - \hat {\b \theta}_{q} \right) \left( \hat{\b \theta}^{(r)}_{q} - \hat {\b \theta}_{q} \right)^{\prime}}.
\end{equation*}
A crucial point when dealing with finite mixtures is the choice of the number of components and/or hidden states. For a fixed quantile level $q$, a simple and frequently used solution is as follows: parameter estimates are computed for varying combinations of the number of components and states, $[G_q,m_q]$, and the model with the best fit, typically measured via log-likelihood or penalized likelihood criteria (such as AIC or BIC), is retained. However, given that these criteria may suffer from early stopping due to lack of progress rather than true convergence and the log-likelihood may present multiple local maxima, the EM algorithm is initialized from different starting points and the model corresponding to the highest likelihood value is retained before applying the penalization. {In this regard, a deterministic start may be employed alongside a set of random start initializations. The deterministic start is obtained by first estimating a linear regression model for the mean response using maximum likelihood, including all covariates -- both those associated with fixed effects and those assumed to have a TC or TV random effect. The estimated fixed regression coefficients serve as initial values for the corresponding parameters.  Initial values for TC and/or TV random coefficients are obtained by adding an appropriate constant to the corresponding fixed effect estimates from the homogeneous model. Prior component probabilities and/or initial probabilities for the latent Markov chain are set uniformly; transition probabilities $\gamma_{k \mid h,q}, k, h = 1, \dots, m$ are instead set to $(1 + w\mathbb I (k = h))/(m+ w)$, where $w$ is a tuning constant.
Random starting points are generated by perturbing the deterministic starting values.}

{While a multi-start strategy may mitigate the risk of local maxima, it is worth highlighting that the final solutions may still lie on the boundary of the parameter space. This can lead to several issues such as (\textit i) splitting components or latent states into multiple subgroups, (\textit{ii}) convergence to (near-)identical parameter estimates across different components or hidden states; (\textit{iii}) inflated standard errors; or (\textit{iv}) instability in fixed-effect estimates. Therefore, it is strongly recommended to carefully examine the selected solution and prefer a simpler model whenever there are indications of the above mentioned issues. To conclude, note that the selection of the optimal number of components and/or states ($G_{q}$ and/or $m_{q}$) should be guided by the need of capturing the unobserved sources of heterogeneity characterizing the data. In this sense, the primary goal is that of providing a sufficiently accurate approximation of the true, possibly continuous, distribution of the random coefficients in the model, rather than identifying homogeneous clusters of  units, as typically done within the finite mixture framework.}

\section[The lqmix  R package]{The \texttt{lqmix}  \textsf{R} package}\label{sec:proposal}

In this section, we introduce the \textsf{R} package \texttt{lqmix}, developed to deal with linear quantile mixture models for longitudinal data. We illustrate the main functions for estimation and inference on the parameters of models described in the previous sections
by considering the application to the labor pain benchmark dataset \citep{Davis1991}. Details on this dataset are given in the following. 

\subsection{Labor pain data}\label{sec:pain}
Firstly reported by \cite{Davis1991} and since then analyzed by \cite{Jung1996} and \cite{GeraciBottai2007} among others, labor pain data come from a randomized clinical trial aiming at analyzing the effectiveness of a medication for relieving labor pain in women. A total of $n=83$ women were randomized to a treatment/placebo group, and a response variable evaluating the self-reported amount of pain measured every $30$ minutes on a 100-$mm$ line was recorded. Here, 0 corresponds to absence of pain, while 100 corresponds to extreme pain. The number of available measurements per woman ranges from a minimum of 1 to a maximum of $T_i = 6$. That is, we are in the presence of an unbalanced longitudinal design, for which a MAR assumption seems to be reasonable. A total of $\sum_{i=1}^{83} T_i = 357$ measurements are available. 

Together with the outcome of interest (\texttt{meas}), two covariates are available: \texttt{trt} denotes an indicator variable identifying the group each woman is assigned to (1 = treatment, 0 = placebo), while \texttt{time} identifies the measurement occasion the response was recorded at. 
Data are stored in the data frame \texttt{pain} included in the \texttt{lqmix} package, as shown below.

\begin{lstlisting}
R> head(pain)
\end{lstlisting}
\begin{lstlisting}
  id meas trt time
1  1  0.0   1    1
2  1  0.0   1    2
3  2  0.0   1    1
4  2  0.0   1    2
5  2  0.0   1    3
6  2  2.5   1    4
\end{lstlisting}

\begin{figure}[h!]
    \begin{center}
         \includegraphics[width=0.65\textwidth]{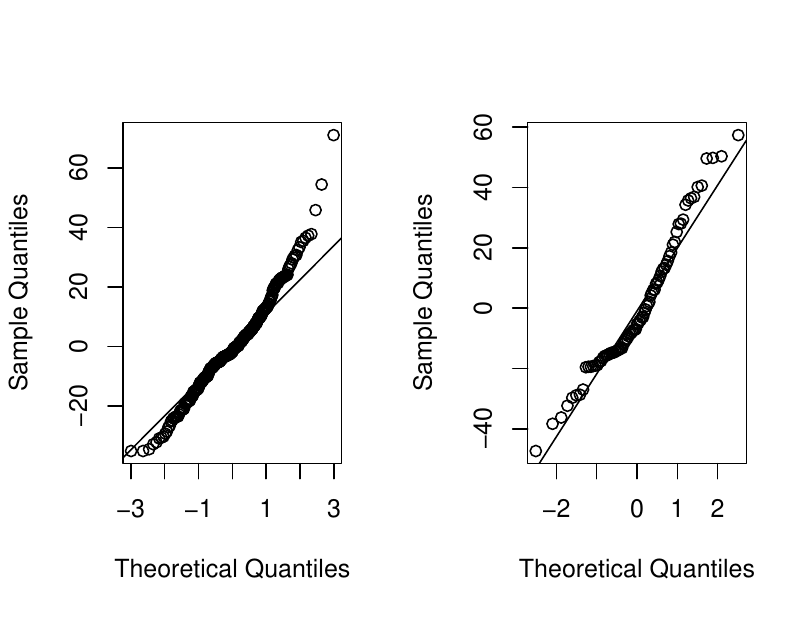}
           \caption{\textcolor{black}{Labor pain data. Linear mixed effect model with Gaussian random intercepts. Normal probability plot for unit- and time-specific residuals (left plot) and for empirical Bayes estimates of unit-specific intercepts (right plot).}} \label{Figure1}
    \end{center}
\end{figure}

Data are severely skewed, and skewness changes magnitude and sign over time. 
In Figure \ref{Figure1}, we report some selected diagnostics for a linear mixed model based on TC, Gaussian, random intercepts, using \texttt{trt}, \texttt{time}, and their interaction (\texttt{trt:time}) as covariates.

\begin{lstlisting}
R> outLME = lmer(meas ~ trt + time + trt:time + (1|id), data = pain)
R> par(mfrow = c(1,2))
R> qqnorm(residuals(outLME), main = "")
R> qqline(residuals(outLME))
R> qqnorm(unlist(ranef(outLME)$id), main = "")
R> qqline(unlist(ranef(outLME)$id))
\end{lstlisting}

In particular, Figure \ref{Figure1} reports the Normal probability plots for model residuals: the left panel shows unit- and time-specific residuals, while the right one shows the empirical Bayes estimates of unit-specific random intercepts. 
As it can be easily noticed, both plots indicate the presence of potentially influential observations in the data, as well as the violation of the Gaussian assumption for the random intercepts in the model. 
Therefore, using linear quantile mixtures seems to be a reasonable choice for the analysis of such data. 

\subsection[The {lqmix} function]{The \texttt{lqmix} function}\label{sec:lqmix}
The main function implemented in the \texttt{lqmix} \textsf{R} package is \texttt{lqmix}. It allows for the estimation  of the parameters of a linear quantile mixture based on either TC, TV, or both types of discrete random  coefficients. The input arguments can be displayed on the \textsf{R} console as follows: 

\begin{lstlisting}[breaklines]
R> args(lqmix)
\end{lstlisting}
\begin{lstlisting}[breaklines]
function (formula, randomTC = NULL, randomTV = NULL, group, time, G = NULL, m = NULL, data, qtl = 0.5, eps = 10^-5, maxit = 1000, se = TRUE, R = 200, start = 0, parInit = list(betaf = NULL, betarTC =  NULL, betarTV = NULL, pg = NULL, delta = NULL, Gamma = NULL, scale = NULL), verbose = TRUE, seed = NULL, parallel = FALSE, ncores = 2) 
\end{lstlisting}
The first mandatory argument, \texttt{formula}, denotes a two-side formula of the type \texttt{(resp $\sim$ Fexpr)}, where \texttt{resp} is the response variable and \texttt{Fexpr} is an expression determining the fixed-coefficient vector, $\b x_{it}$. On the other side, \texttt{randomTC} and \texttt{randomTV} are two one-side formulas of the type \texttt{($\sim$ Rexpr1)} and \texttt{($\sim$ Rexpr2)}, where \texttt{Rexpr1} and \texttt{Rexpr2} identify the columns of covariates associated with TC and TV random-coefficient vectors, $\b z_{it}$ and $\b w_{it}$, respectively. Note that both these arguments are optional, so that the user may decide to estimate a linear quantile regression model with either TC or TV random coefficients, or with a combination of them. Further, variables reported in \texttt{Rexpr1} and \texttt{Rexpr2} must not overlap and, if any of them appears also in the fixed effect formula (\texttt{formula}), the corresponding parameter is estimated as a TC/TV random coefficient only. The arguments \texttt{group} and \texttt{time} are strings indicating the grouping and time variable, respectively. All such variables are taken from the data frame specified through the  (mandatory) argument \texttt{data}.
The arguments \texttt{G} and \texttt{m} are used to specify the number of mixture components and/or hidden states in the model, respectively, while \texttt{qtl} allows to specify the quantile level (by default, \texttt{qtl = 0.5}). The arguments \texttt{se}, \texttt{R},  and \texttt{start} allow to specify whether block-bootstrap standard errors should be computed (by default, \texttt{se = TRUE}), the number of bootstrap samples to be used for this purpose ({by default, \texttt{R = 200}}), and the initialization rule to consider. As regards this latter, three possible specifications are allowed: \texttt{start = 0} is used for a deterministic start of model parameters (the default option); 
\texttt{start = 1} is used for a random start of model parameters; \texttt{start = 2} is used to consider a starting rule based on given model parameters specified via the \texttt{parInit} list argument. {The arguments} \texttt{maxit}, \texttt{eps}, and \texttt{verbose} identify the maximum number of iterations for the EM algorithm (by default, \texttt{maxit = 1000}),  the corresponding tolerance level (by default  \texttt{eps = $\mathtt{10^{-5}}$}), and whether output should be printed (by default, \texttt{verbose = TRUE}), {respectively}. As far as the  \texttt{seed} argument, this is devoted to setting a seed for random number generation that is used for the random starting rule (\texttt{start=1}) described so far and computing model parameters' standard errors. The arguments \texttt{parallel} and \texttt{ncores} control parallel computing when standard errors are requested, with \texttt{ncores} defaulting to 2.

\subsubsection{Estimating linear quantile mixtures with TC random coefficients}\label{sec:lqmixTC}
To explore features of the \texttt{lqmix} function, we start by considering a two-component linear quantile mixture ($\mathtt{G_{q} = 2}$) for the median ($\mathtt{qtl = 0.5}$) with a TC random intercept for the analysis of the pain data. This is estimated as follows: 

\begin{lstlisting}[breaklines]
R> outTC = lqmix(formula = meas ~ time + trt + trt:time, randomTC = ~1, time = "time", group = "id", G = 2, data = pain)
\end{lstlisting}
\begin{lstlisting}[breaklines]
--------|-------|-------|--------|-------------|-------------|
  model |  qtl  |   G   |  iter  |      lk     |   (lk-lko)  |
--------|-------|-------|--------|-------------|-------------|
     TC |   0.5 |     2 |      0 |    -1682.31 |          NA | 
     TC |   0.5 |     2 |      8 |    -1607.11 | 4.64719e-06 | 
--------|-------|-------|--------|-------------|-------------|
Computing standard errors ...
  |==============================================================| 100%
  \end{lstlisting}

The running time for the above command is 1.930 seconds when run on an Apple M1 architecture (16GB, MacOS: Sequoia 15.5). This represents the architecture used for all the codes illustrated in the paper. 
In the following, we report the output of the above call to \texttt{lqmix}, which is an object of class \texttt{lqmix}, obtained by using the \texttt{print} method of the \texttt{S3} class. 
\begin{lstlisting}[breaklines]
R> outTC
\end{lstlisting}
\begin{lstlisting}[breaklines]
Model: TC random coefficients with G=2 at qtl=0.5
****************************************************** 

---- Observed process ----

Fixed Coefficients:
    time      trt time.trt 
 11.6669  -2.1659 -10.8335 

Time-Constant Random Coefficients:
      (Intercept)
Comp1      1.3325
Comp2     43.3325

Residual scale parameter: 7.3289 - Residual standard deviation: 20.7294 

---- Latent process ----

Mixture probabilities:
 Comp1  Comp2 
0.6737 0.3263 

Log-likelihood at convergence: -1607.11
Number of observations: 357 - Number of subjects: 83 
\end{lstlisting}
Looking at the output, we may recognize two separate sections, reporting estimates for the observed and the latent process, respectively. 
As regards the observed process, the estimated fixed coefficients suggest how self-reported pain increases as the time passes by, together with a positive effect of the treatment under investigation (negative sign for the \texttt{trt} variable), with benefits that increase with time.
On the other side, the estimated random coefficients highlight the presence of two well-separated groups of women, reporting a low and a medium pain at the baseline, respectively. In the last part, the estimated scale parameter and the corresponding error standard deviation are shown. This latter corresponds to the standard deviation of an ALD.
As regards the latent process, estimates highlight that the 67.37\% of women belong to the first mixture component (low-pain level); the remaining 32.63\% belong to the second one (medium-pain level). 
All estimated parameters are stored in the object \texttt{outTC} as \texttt{betaf} (the fixed coefficients), \texttt{betarTC} (the TC random coefficients), \texttt{scale} (the scale parameter), \texttt{sigma.e} (the conditional standard deviation of responses),
and \texttt{pg} (the component prior probabilities). Further, when leaving the statement \texttt{se = TRUE} (default value), the   \texttt{outTC} object also contains information on the estimated standard errors of model parameters obtained via the bootstrap procedure detailed in Section ``\textit{Standard Errors and model selection}''. These standard errors can be accessed by prefixing the corresponding parameter names with \texttt{se.}, as listed above.

The following information are also stored in the the \texttt{outTC} object: the log-likelihood value at convergence (\texttt{lk}), the number of model parameters (\texttt{npar}),  the values of AIC and BIC (\texttt{aic} and \texttt{bic}), the quantile level (\texttt{qtl}), the number of mixture components (\texttt{G}), the total number of subjects and observations (\texttt{nsbjs} and \texttt{nobs}), the mixture components' posterior probabilities obtained at convergence of the EM algorithm (\texttt{postTC}), the bootstrap variance-covariance matrices of the regression coefficients (\texttt{vcov}), the type of missingness data are affected by (\texttt{miss}), the estimated model (\texttt{model}), and the model call (\texttt{call}). Last, model matrices associated with fixed and TC random coefficients are stored in the \texttt{outTC} object as \texttt{mmf} and \texttt{mmrTC}, respectively.

\subsubsection{Estimating linear quantile mixtures with TV random coefficients}\label{sec:lqmixTV}
The same \texttt{lqmix} function can be used to obtain parameter estimates for a quantile mixture with TV random coefficients. To analyze the labor pain data, we consider a linear quantile mixture for the median ($\mathtt{qtl = 0.5}$), with a TV random intercept defined over a two-state latent space ($\mathtt{m_{q} = 2}$): 
\begin{lstlisting}[breaklines]
R> outTV = lqmix(formula = meas ~ time + trt + trt:time, randomTV = ~1, time = "time", group = "id", m = 2, data = pain)
\end{lstlisting}
\begin{lstlisting}[breaklines]
--------|-------|-------|--------|-------------|-------------|
  model |  qtl  |   G   |  iter  |      lk     |   (lk-lko)  |
--------|-------|-------|--------|-------------|-------------|
     TV |   0.5 |     2 |      0 |    -1688.99 |          NA | 
     TV |   0.5 |     2 |     10 |    -1576.61 |    0.366345 | 
     TV |   0.5 |     2 |     20 |    -1575.79 | 9.32048e-05 | 
     TV |   0.5 |     2 |     29 |    -1575.79 | 9.43243e-06 | 
--------|-------|-------|--------|-------------|-------------|
Computing standard errors ...
  |==============================================================| 100% 
\end{lstlisting}
The running time for obtaining results is 3.351  seconds. The output of estimation is given below. 
\begin{lstlisting}[breaklines]
R> outTV
\end{lstlisting}
\begin{lstlisting}[breaklines]
Model: TV random coefficients with m=2 at qtl=0.5
****************************************************** 

---- Observed process ----

Fixed Coefficients:
    time      trt time.trt 
  6.5012  -0.4923  -6.0013 

Time-Varying Random Coefficients:
    (Intercept)
St1      0.4924
St2     60.9926

Residual scale parameter: 5.8337 - Residual standard deviation: 16.5003 

---- Latent process ----

Initial probabilities:
   St1    St2 
0.7667 0.2333 

Transition probabilities:
         toSt1  toSt2
fromSt1 0.8883 0.1117
fromSt2 0.0271 0.9729

Log-likelihood at convergence: -1575.795
Number of observations: 357 - Number of subjects: 83 
\end{lstlisting}
Results for the observed process allow to derive similar conclusions to those detailed in the previous section, even though the magnitude of the effects is reduced. 
As regards the latent layer of the model, the \texttt{print} method shows the estimates for the initial and the transition probabilities of the hidden Markov chain. Based on such estimates, one may conclude that 76.67\% of women start in the first hidden state which,  based on the estimated time-varying random intercepts, is characterized by a low pain level. The remaining 23.33\% of women start the study with an intermediate baseline pain level. Looking at the estimated transition probabilities, we may notice that, regardless of the treatment effect, labor pain tends to increase as the time passes by, with women being in the \textit{low pain} state moving towards the \textit{medium pain} state with probability equal to $0.1117$.
Estimated parameters are stored in the object \texttt{outTV} as \texttt{betaf} (the fixed coefficients), \texttt{betarTV} (the TV random coefficients), \texttt{scale} (the scale parameter), \texttt{sigma.e} (the conditional standard deviation of responses derived from an ALD with parameters \texttt{scale} and \texttt{qtl}), \texttt{delta} (the initial probability vector), and \texttt{Gamma} (the transition probability matrix). As detailed in the previous section, bootstrap standard errors are stored by using the prefix ``\texttt{se.}'' to the names of parameters of interest. Last, posterior probabilities for the latent states of the hidden Markov chain and the model matrix for the TV random coefficients are stored as \texttt{posTV} and \texttt{mmrTV}, respectively, in the \texttt{outTV} object.
 
\subsubsection{Estimating linear quantile mixtures with TC and TV random coefficients} \label{sec:lqmix_TCTV}
When specifying both the \texttt{randomTC} and the \texttt{randomTV} formulas, and therefore both \texttt{G} and \texttt{m}, the \texttt{lqmix} function allows for the estimation of linear quantile mixtures based on both TC and TV, discrete, random coefficients. 
To analyze labor pain data, we consider a linear quantile mixture for the median ({\tt qtl = 0.5}), with a TV random intercept, with $m_{q} = 2$ hidden states, and a TC random slope for the \texttt{time} variable, with $G_{q} = 2$ components. To estimate such a model and consider a random start initialization (\texttt{start=1}), the following command can be run: 
\begin{lstlisting}[breaklines]
R> outTCTV = lqmix(formula = meas ~ trt + time + trt:time, randomTC = ~ time, randomTV = ~1, time = "time", group = "id", m = 2, G = 2, data = pain, se = FALSE, start = 1, seed = 10)
\end{lstlisting}
\begin{lstlisting}[breaklines]
--------|-------|-------|-------|--------|-------------|-------------|
  model |  qtl  |   m   |   G   |  iter  |      lk     |   (lk-lko)  |
--------|-------|-------|-------|--------|-------------|-------------|
   TCTV |   0.5 |     2 |     2 |      0 |    -1723.86 |          NA | 
   TCTV |   0.5 |     2 |     2 |     10 |    -1594.61 |     14.8135 | 
   TCTV |   0.5 |     2 |     2 |     20 |    -1554.41 |     3.47302 | 
   TCTV |   0.5 |     2 |     2 |     30 |    -1541.16 |    0.165982 | 
   TCTV |   0.5 |     2 |     2 |     37 |    -1541.12 | 2.98304e-06 | 
--------|-------|-------|-------|--------|-------------|-------------|
\end{lstlisting}
This requires a running time of 0.240 seconds. 
The output of the above call to \texttt{lqmix}, obtained through the \texttt{print} method of the \texttt{S3} class, is reported below.
\begin{lstlisting}
R> outTCTV
\end{lstlisting}
\begin{lstlisting}
Model: TC and TV random coefficients with m=2 and G=2 at qtl=0.5
****************************************************************** 

---- Observed process ----

Fixed Coefficients:
     trt trt.time 
 -9.2859  -5.0428 

Time-Constant Random Coefficients:
         time
Comp1  5.5428
Comp2 14.9178

Time-Varying Random Coefficients:
    (Intercept)
St1      8.7859
St2     69.7859

Residual scale parameter: 5.1259 - Residual standard deviation: 14.4983 

---- Latent process ----

Mixture probabilities:
 Comp1  Comp2 
0.7381 0.2619 

Initial probabilities:
   St1    St2 
0.8036 0.1964 

Transition probabilities:
         toSt1  toSt2
fromSt1 0.9641 0.0359
fromSt2 0.0434 0.9566

Log-likelihood at convergence: -1541.12
Number of observations: 357 - Number of subjects: 83 
\end{lstlisting}

As it is clear from the output, the linear quantile mixture with both TC and TV, discrete, random coefficients produces more refined information when compared to those discussed so far.
Also in this case, we may recognize in the output the two sections entailing the observed and the latent process, respectively. In the former, estimates for the fixed, the TC, and the TV random coefficients are reported. 
For the latent part, the output reports information on the mixture, the initial, and the transition probabilities. 

By looking at the results, we conclude again that self reported pain is lower for women under medication (negative sign for the \texttt{trt} variable) and that  benefits increase with time. Looking at the estimated TC random coefficients and the corresponding prior probabilities, we may distinguish  two well separated groups of women.  They exhibit different trends in pain levels over time: 73.81\% experience a mild increase in pain levels, whereas 26.19\% show a steeper increase. 
On the other side, the estimated TV random coefficients identify two groups of women characterized by a low and a medium baseline labor pain level, respectively. At the beginning of the observation period, the 80.36\% of women belong to the first group, while the remaining 19.64\% to the second. By looking at the estimated transition probability matrix,  we conclude that the group composition remains largely unchanged over time once controlling for the other effects in the model ($\gamma_{hh}>0.95, h = 1, 2$). 

Estimated parameters are stored in the object \texttt{outTCTV}. Now both those related to the TC finite mixture as well as those related to the hidden Markov chain are referenced.
 In detail, \texttt{outTCTV} contains \texttt{betaf} (the fixed coefficients), \texttt{betarTC} (the TC random coefficients), \texttt{betarTV} (the TV random coefficients), \texttt{scale} (the scale parameter), \texttt{sigma.e} (the conditional standard deviation of responses derived from an ALD with parameters \texttt{scale} and \texttt{qtl}), \texttt{pg} (the component prior probabilities), \texttt{delta} (the initial probabilities), and \texttt{Gamma} (the transition probabilities). Standard errors (if computed), as well as additional information on the data, the estimated model, posterior probabilities, and model matrices are stored in the \texttt{outTCTV} object.

\subsection[The {searchlqmix} function for model selection]{The \texttt{search\_lqmix} function for model selection}
As described in Section ``\textit{Standard errors and model selection}'', the number of mixture components $G_q$ and the number of hidden states $m_q$ in the model are unknown quantities that need to be estimated. Moreover, a multi-start strategy is frequently needed to solve, at least partially,  the potential multimodality of the likelihood surface. Both issues can be addressed by means of the \texttt{search\_lqmix()} function.
Input arguments and corresponding default values can be displayed on the \textsf{R} console though the  \texttt{args} function:
\begin{lstlisting}[breaklines]
R> args(search_lqmix)
\end{lstlisting}
\begin{lstlisting}[breaklines]
function (formula, randomTC = NULL, randomTV = NULL, group, time, Gv = NULL, mv = NULL, data, method = "bic", nran = 0, qtl = 0.5, eps = 10^-5, maxit = 1000, se = TRUE, R = 200, verbose = TRUE, seed = NULL, parallel = FALSE, ncores = 2) 
\end{lstlisting}
Most of the arguments of this function correspond to those described so far. 
The remaining ones are specified as follows. \texttt{Gv} and \texttt{mv} denote vectors identifying the range for the number of mixture components, $G_q$, and the number of hidden states, $m_q$, to be considered, respectively, for a fixed quantile level $q \in (0,1)$. When both arguments are specified, the search of the optimal linear quantile mixture with TC and TV random coefficients is performed. When only one out the two arguments is specified, a linear quantile mixture based on either TC or TV random coefficients is estimated. A linear quantile regression with no random coefficients is also estimated either when both \texttt{mv} and/or \texttt{Gv} include the value 1, or when \texttt{Gv} includes 1 and \texttt{mv = NULL}, or when \texttt{mv} includes 1 and \texttt{Gv = NULL}. In this case, the function \texttt{lqr()} implemented in the \texttt{lqmix} \textsf{R} package and described in the following is employed.  

The argument \texttt{method} is used for model selection purposes.  One of three possible values is admitted: ``\texttt{bic}'' (by default), ``\texttt{aic}'', or ``\texttt{lk}''. The former two identify the optimal model as that providing the minimum value of the BIC or the AIC index, respectively. The latter, selects the optimal model as the one corresponding to the maximum log-likelihood value. 

The argument \texttt{nran} specifies the number of random starts to be considered for each value in the range identified by  \texttt{Gv} and \texttt{mv}. This number is strictly related to the type of model for which the optimal search is performed. In detail, following a similar strategy as that suggested by \cite{LMestPack} for the \texttt{LMest} \textsf{R} package, when a linear quantile mixture based on TC random coefficients only is considered, the number of random initializations is set equal to $\mathtt{nran\times(G_q-1)}$; when only TV random coefficients are allowed, the number of random initializations is set equal to $\mathtt{nran\times(m_q-1)}$; last, when both TC and TV random coefficients are considered, the number of random initializations is set equal to $\mathtt{nran\times(G_q-1)\times(m_q-1)}$. By default, $\mathtt{nran = 0}$, so that no random initializations are considered. The \texttt{seed} argument is used to fix a seed and ensure reproducibility of results when the multi-start strategy based on random initializations is considered to estimate model parameters, as well as for deriving standard errors (when requested). Last, \texttt{parallel} and \texttt{ncores} are used for parallel computing of the standard errors. 

We report below the results of the model selection strategy applied to the \texttt{pain} data, when focusing on a linear quantile mixture with a TV random intercept and a TC random slope associated to the variable \texttt{time}, for the quantile level $q = 0.5$. The search is done by looking for the optimal number of components ($G_{q}$) and hidden states ($m_{q}$), both in the set $\{1,2 \}$, by setting \texttt{nran=50}. The optimal model is selected according to the BIC index.
\begin{lstlisting}[breaklines]
R> sTCTV = search_lqmix(formula = meas ~ trt + time + trt:time, randomTC = ~time, randomTV = ~1, group = "id", time = "time", nran = 50, mv = 1:2, Gv = 1:2, data = pain, seed = 10)
\end{lstlisting}
\begin{lstlisting}[breaklines]
Search the optimal linear quantile mixture model 
************************************************* 
Random start: 0 ... 
--------|-------|--------|-------------|
  model |  qtl  |  iter  |      lk     |
--------|-------|--------|-------------|
    HOM |   0.5 |      0 |    -1707.73 | 
--------|-------|--------|-------------|
Random start: 0 ... 1  ... 2  ... 3  ... 4  ... 5  ... 6  ... 7  ... 8  ... 9  ... 10  ... 11  ... 12  ... 13  ... 14  ... 15  ... 16  ... 17  ... 18  ... 19  ... 20  ... 21  ... 22  ... 23  ... 24  ... 25  ... 26  ... 27  ... 28  ... 29  ... 30  ... 31  ... 32  ... 33  ... 34  ... 35  ... 36  ... 37  ... 38  ... 39  ... 40  ... 41  ... 42  ... 43  ... 44  ... 45  ... 46  ... 47  ... 48  ... 49  ... 50  ... 
--------|-------|-------|--------|-------------|-------------|
  model |  qtl  |   G   |  iter  |      lk     |   (lk-lko)  |
--------|-------|-------|--------|-------------|-------------|
     TC |   0.5 |     2 |      0 |    -1626.59 |          NA | 
     TC |   0.5 |     2 |      1 |    -1626.59 | 6.65801e-07 | 
--------|-------|-------|--------|-------------|-------------|
Random start: 0 ... 1  ... 2  ... 3  ... 4  ... 5  ... 6  ... 7  ... 8  ... 9  ... 10  ... 11  ... 12  ... 13  ... 14  ... 15  ... 16  ... 17  ... 18  ... 19  ... 20  ... 21  ... 22  ... 23  ... 24  ... 25  ... 26  ... 27  ... 28  ... 29  ... 30  ... 31  ... 32  ... 33  ... 34  ... 35  ... 36  ... 37  ... 38  ... 39  ... 40  ... 41  ... 42  ... 43  ... 44  ... 45  ... 46  ... 47  ... 48  ... 49  ... 50  ... 
--------|-------|-------|--------|-------------|-------------|
  model |  qtl  |   G   |  iter  |      lk     |   (lk-lko)  |
--------|-------|-------|--------|-------------|-------------|
     TV |   0.5 |     2 |      0 |    -1575.79 |          NA | 
     TV |   0.5 |     2 |      4 |    -1575.79 | 8.79175e-06 | 
--------|-------|-------|--------|-------------|-------------|
Random start: 0 ... 1  ... 2  ... 3  ... 4  ... 5  ... 6  ... 7  ... 8  ... 9  ... 10  ... 11  ... 12  ... 13  ... 14  ... 15  ... 16  ... 17  ... 18  ... 19  ... 20  ... 21  ... 22  ... 23  ... 24  ... 25  ... 26  ... 27  ... 28  ... 29  ... 30  ... 31  ... 32  ... 33  ... 34  ... 35  ... 36  ... 37  ... 38  ... 39  ... 40  ... 41  ... 42  ... 43  ... 44  ... 45  ... 46  ... 47  ... 48  ... 49  ... 50  ... 
--------|-------|-------|-------|--------|-------------|-------------|
  model |  qtl  |   m   |   G   |  iter  |      lk     |   (lk-lko)  |
--------|-------|-------|-------|--------|-------------|-------------|
   TCTV |   0.5 |     2 |     2 |      0 |    -1538.76 |          NA | 
   TCTV |   0.5 |     2 |     2 |      3 |    -1538.76 | 8.36816e-06 | 
--------|-------|-------|-------|--------|-------------|-------------|
Computing standard errors for the optimal model...
  |==============================================================| 100%
\end{lstlisting}
The running time for the above command is 20.942 seconds and the output is stored in the \texttt{sTCTV} object. This can be shown by means of the \texttt{print} method of the class \texttt{S3} as follows:
\begin{lstlisting}[breaklines]
R>  sTCTV
\end{lstlisting}
\begin{lstlisting}[breaklines]
Opt model: TC and TV random coefficients with m=2 and G=2 at qtl=0.5
********************************************************************* 

---- Observed process ----

Fixed Coefficients:
     trt trt.time 
 -4.3237  -5.0294 

Time-Constant Random Coefficients:
         time
Comp1  5.3294
Comp2 15.6627

Time-Varying Random Coefficients:
    (Intercept)
St1      4.0237
St2     49.1746

Residual scale parameter: 4.7858 - Residual standard deviation: 13.5363 

---- Latent process ----

Mixture probabilities:
 Comp1  Comp2 
0.6351 0.3649 

Initial probabilities:
   St1    St2 
0.7553 0.2447 

Transition probabilities:
         toSt1  toSt2
fromSt1 0.9552 0.0448
fromSt2 0.1497 0.8503

Log-likelihood at convergence: -1538.76
Number of observations: 357 - Number of subjects: 83 
\end{lstlisting}
As reported in the first output line, the optimal model for the median (\texttt{qtl = 0.50}) according to the BIC criterion (the default option) is based on $G_{q} = 2$ mixture components and $m_{q} = 2$ hidden states. Results we obtain are in line with those reported in the previous section. Differences are due to the model selection strategy that aims at identifying the global maximum of the likelihood function. 

{To provide further insights into the analysis of the \texttt{pain} data and look at the potential of the \texttt{lqmix R} package, we also search for the optimal model for a different quantile level. Specifically, we apply the \texttt{search\_lqmix} function for \texttt{qtl = 0.75} to study the impact of  observed covariates on higher pain levels. When looking at the results reported below, we may notice that again the optimal specification (according to the BIC criterion) is obtained for $G_{q}=2$ and $m_{q} = 2$. 
Further, pain levels are lower when medication is taken, even though the estimated effect  is lower than before. This results claims at a lower beneficial effect of treatment for those women reporting higher pain levels. Comparing the estimated TC random coefficients for \texttt{qlt = 0.75} with those from the model for \texttt{qtl = 0.50}, we observe now a more pronounced increase of pain levels over time; the estimated TV random coefficients identify instead two groups of women declaring again a low and a medium baseline pain level, respectively. Baseline estimates, as expected, are now higher than those obtained for \texttt{qtl = 0.50}. The estimated transition probability matrix highlights a very high persistence in each of the two states; this means that baseline levels remain pretty constant over time.
}
\begin{lstlisting}[breaklines]
R> sTCTV75 = search_lqmix(formula = meas ~ trt + time + trt:time, randomTC = ~time, randomTV = ~1, nran = 50, group = "id", time =     		"time", mv = 1:2, Gv = 1:2, data = pain, seed = 10, qtl = 0.75)
\end{lstlisting}
\begin{lstlisting}[breaklines]
R> sTCTV75
\end{lstlisting}
\begin{lstlisting}
Opt model: TC and TV random coefficients with m=2 and G=2 at qtl=0.75
********************************************************************* 

---- Observed process ----

Fixed Coefficients:
     trt trt.time 
 -1.0538  -7.6027 

Time-Constant Random Coefficients:
         time
Comp1  8.5027
Comp2 17.9231

Time-Varying Random Coefficients:
    (Intercept)
St1      5.1538
St2     65.9891

Residual scale parameter: 4.2329 - Residual standard deviation: 17.8476 

---- Latent process ----

Mixture probabilities:
 Comp1  Comp2 
0.6965 0.3035 

Initial probabilities:
   St1    St2 
0.7393 0.2607 

Transition probabilities:
         toSt1  toSt2
fromSt1 0.9761 0.0239
fromSt2 0.0526 0.9474

Log-likelihood at convergence: -1577.189
Number of observations: 357 - Number of subjects: 83
\end{lstlisting}

\subsection[The {summary} method for {lqmix} and search\_lqmix objects]{The \texttt{summary} method for \texttt{lqmix} and \texttt{search\_lqmix} objects}\label{sec:summary}
The \texttt{summary} method, when applied to objects of class \texttt{lqmix} or \texttt{search\_lqmix}, generates a summary object that allows for inference on model parameters. For \texttt{search\_lqmix} objects, it returns the summary of the optimal model. In the following, we present the output of the \texttt{summary} method for the \texttt{sTCTV} object described so far.
\begin{lstlisting}
R> summary(sTCTV)
\end{lstlisting}
\begin{verbatim}
Opt model: TC and TV random coefficients with m=2 and G=2 at qtl=0.5
*************************************************************************** 

---- Observed process ----

Fixed Coefficients:
         Estimate St.Error z.value P(>|z|)    
trt       -4.3237   2.6423 -1.6363  0.0997 .  
trt.time  -5.0294   0.7494 -6.7108  <2e-16 ***

Time-Constant Random Coefficients:
           Estimate St.Error z.value   P(>|z|)    
time_Comp1   5.3294   0.5207  10.236 < 2.2e-16 ***
time_Comp2  15.6627   0.8050  19.457 < 2.2e-16 ***

Time-Varying Random Coefficients:
                Estimate St.Error z.value P(>|z|)    
(Intercept)_St1   4.0237   1.6985   2.369  0.0175 *  
(Intercept)_St2  49.1746   4.8451  10.149  <2e-16 ***
---
Signif. codes:  0 ‘***’ 0.001 ‘**’ 0.01 ‘*’ 0.05 ‘.’ 0.1 ‘ ’ 1

Residual scale parameter: 4.7858 - Residual standard deviation: 13.5363 

---- Latent process ----

Mixture probabilities:
      Estimate St.Error
Comp1   0.6351   0.0668
Comp2   0.3649   0.0668

Initial probabilities:
    Estimate St.Error
St1   0.7553   0.0507
St2   0.2447   0.0507

Transition probabilities:
             Estimate St.Error
fromSt1toSt1   0.9552   0.0181
fromSt1toSt2   0.0448   0.0181
fromSt2toSt1   0.1497   0.0546
fromSt2toSt2   0.8503   0.0546

Log-likelihood at convergence: -1538.76
Number of observations: 357 - Number of subjects: 83 
\end{verbatim}
As before, two different sections may be distinguished. In the former, estimates, standard errors, test statistics, and corresponding approximate p-values for assessing significance of model parameters for the observed process the are reported. These information are completed with the estimates of the scale parameter and the corresponding conditional standard deviation of the error terms. 
In the latter section of the output, estimates and standard errors for the parameters characterizing the latent layer of the model are reported.

\section[Other methods for lqmix, searchlqmix, and lqr objects]{Other methods for \texttt{lqmix} and \texttt{search\_lqmix} objects}
\begin{figure}
\caption{Results from the \texttt{plot} method}
\centering 
\includegraphics[scale = 0.38, page = 1]{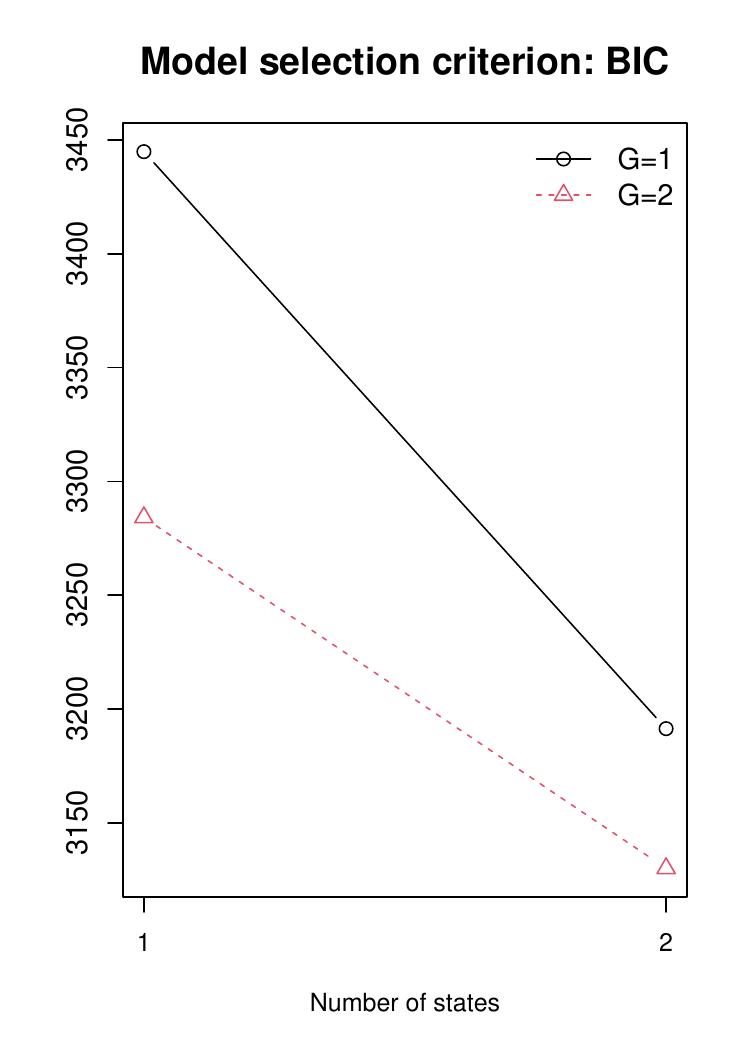}
\hspace{-5mm}\includegraphics[scale = 0.38, page = 2]{figplot.pdf}
\hspace{-9mm}\includegraphics[scale = 0.38, page = 3]{figplot.pdf}
\label{fig:plot}
\end{figure}

To provide users with a familiar and consistent interface, the \texttt{lqmix} \textsf{R} package includes basic methods typically available for regression model objects. Together with the \texttt{print} and the \texttt{summary} methods introduced in the previous sections, the \texttt{logLik} and \texttt{coef} methods are implemented. These return the maximum log-likelihood value obtained at convergence of the EM algorithm and the estimated fixed effect coefficients, respectively. Further, the methods \texttt{AIC} and \texttt{BIC} may be used for model selection purposes. When the above methods are applied to objects of class \texttt{search\_lqmix}, these refer to the optimal specification identified via the \texttt{search\_lqmix} function. The \texttt{plot} method returns graphical representations for the outputs of the \texttt{lqmix} and the \texttt{search\_lqmix} functions. In detail, a graphical display of the mixture component probabilities and/or of the transitions across states of the hidden Markov chain is provided when applied to an \texttt{lqmix} object.
 A plot reporting the value of the chosen model selection criterion for varying number of components and/or hidden states (defined in \texttt{Gv} and \texttt{mv}, respectively) is also provided when applied to objects of class \texttt{search\_lqmix}. {We report in Figure \ref{fig:plot} the results of such a method applied to the \texttt{sTCTV} object detailed in Section ``\textit{The \texttt{search\_lqmix} function for model selection}''. 
 \\\indent
Further methods implemented in the package are \texttt{predict}, \texttt{residuals}, and \texttt{vcov}. The former two return the conditional predicted values and the conditional residuals, given the component membership and/or the state occupied at a given occasion by units in the sample; \texttt{vcov} returns instead  the variance-covariance matrix of fixed model parameters obtained from the block-bootstrap procedure detailed above.  } 

\section{Linear quantile regression for independent data}
The package \texttt{lqmix} is thought for dealing with longitudinal data. However, it also implements the function \texttt{lqr()} for estimating a homogeneous linear quantile regression model via a ML approach. This is obtained by exploiting the parallelism between the ALD and the asymmetric quantile loss function. 
Input arguments for the \texttt{lqr} function and corresponding default values are as follows: 
\begin{lstlisting}
R> args(lqr)
\end{lstlisting}
\begin{lstlisting}[breaklines]
function (formula, data, qtl = 0.5, se = TRUE, R = 100, verbose = TRUE, seed = NULL, parallel = FALSE, ncores = 2) 
\end{lstlisting}

These arguments are identical to those described above for the functions that are built to estimate parameters of linear quantile mixtures. As evident, also in this case the user may specify whether standard errors need to be computed. In this case, a block-bootstrap approach is employed. By defaults, \texttt{R = 200}  re-samples are considered. 
{All methods described in the previous section and available for \texttt{lqmix} and \texttt{search\_lqmix} objects are also available for for object of class \texttt{lqr}.}

\section{Conclusions}

Quantile regression is a fundamental tool of analysis when (\textit{i}) the response distribution is skewed, (\textit{ii}) outlying values are present, (\textit{iii})  interest entails both the center and the tails of the response distribution. When dealing with longitudinal data, dependence between observations from the same subject must be properly considered to avoid biased inference. Thus, linear quantile regression models with unit-specific random coefficients  may be effectively employed. 

In this paper, we present the \texttt{lqmix}  \textsf{R} package to estimate mixtures of linear quantile regression models for continuous longitudinal data, possibly affected by random missingness. Strong and unverifiable parametric assumptions on the random coefficients are avoided by considering a mixture specification. Either TC or TV random coefficients, or a combination of both can be fitted to deal with different sources of unobserved heterogeneity affecting the longitudinal process. That is, the package allows us to deal with unit-specific unobserved features (omitted covariates in the model) which may vary and/or stay constant over the observational period, while looking at the quantiles of the conditional response variable. 

{Three main functions are implemented in the package. A first one, \texttt{lqmix}, allows the estimation of the different model specifications obtained by considering TC or TV random coefficients only, or a combination of them. A second function, \texttt{search\_lqmix}, allows for the searching of the optimal model specification, based on various model selection criteria. A last function, \texttt{lqr}, is devoted to the estimation of linear quantile regression models for cross-sectional data. For a concise understanding, we report in Table \ref{tab:summary} the description of the main parameters required by such functions. }
\begin{table}[htbp]
{

  \caption{Description of the main arguments for the functions \texttt{lqmix}, \texttt{search\_lqmix}, and \texttt{lqr}.}
  
\scalebox{0.85}{
\centering
    \begin{tabular}{llccc}
    \toprule
    Arguments & Description & \texttt{lqmix} & \texttt{search\_lqmix} & \texttt{lqr} \\
    \midrule
    \texttt{formula} & \multicolumn{1}{p{25em}}{an object of class \texttt{formula}: a symbolic description of the model to be fitted of the form \texttt{resp $\sim$ Fexpr}} & yes   & yes   & yes \\
    \texttt{randomTC} & \multicolumn{1}{p{25em}}{a one-sided formula of the form $\sim z_{1}+z_{2}+...+z_{r}$} & yes   & yes    & no \\
    \texttt{randomTV} & \multicolumn{1}{p{25em}}{a one-sided formula of the form $\sim w_{1}+w_{2}+...+w_{l}$} & yes   & yes    & no \\
    \texttt{group} & \multicolumn{1}{p{25em}}{a string indicating the grouping variable} & yes   & yes   & no \\
    \texttt{time} & \multicolumn{1}{p{25em}}{a string indicating the time variable} & yes   & yes   & no \\
    \texttt{G} & \multicolumn{1}{p{25em}}{number of mixture components associated to TC random coefficients} & yes   & no    & no \\
    \texttt{m} & \multicolumn{1}{p{25em}}{number of states associated to the TV random coefficients} & yes   & no    & no \\
    \texttt{Gv} & \multicolumn{1}{p{25em}}{vector of possible number of mixture components associated to TC random coefficients } & no    & yes   & no \\
    \texttt{mv} & \multicolumn{1}{p{25em}}{vector of possible number of mixture components associated to TV random coefficients} & no    & yes   & no \\
    \texttt{data} & \multicolumn{1}{p{25em}}{a data frame containing the variables named in \texttt{formula}, \texttt{randomTC}, \texttt{randomTV}, \texttt{group}, and \texttt{time}} & yes   & yes   & yes \\
    \texttt{qtl} & quantile to be estimated & yes   & yes   & yes \\
    \texttt{eps} &  \multicolumn{1}{p{25em}}{tolerance level for (relative) convergence of the EM algorithm }& yes   & yes   & no \\
    \texttt{maxit} & \multicolumn{1}{p{25em}}{maximum number of iterations for the EM algorithm} & yes   & yes   & no \\
    \texttt{se} & \multicolumn{1}{p{25em}}{standard error computation for the optimal model} & yes   & yes   & yes \\
    \texttt{R} & \multicolumn{1}{p{25em}}{number of bootstrap samples for computing standard errors} & yes   & yes   & yes \\
    \texttt{nran} & \multicolumn{1}{p{25em}}{number of repetitions of each random initialization} & no    & yes   & no \\
    \bottomrule
    \end{tabular}%
}
  \label{tab:summary}%
}
\end{table}%

{When looking at the scalability of \texttt{lqmix}, the computational effort increases with the complexity of the model. Linear quantile mixtures of quantile regressions with TC random coefficients are computationally less demanding than those with TV random coefficients, which in turn are more efficient than models incorporating both TC and TV random terms. Additional factors influencing computational load include the number of repeated measures and, most significantly, the number of units in the sample.
In our experience, \texttt{lqmix} easily handles datasets of small to moderate size (about 10 thousand units). For larger datasets, the computational effort increases, particularly in the estimation of standard errors. In this respect, parallel computing proves beneficial in managing computation times and maintaining performance.}

{Another aspect that is important to highlight entails the selection of the number of bootstrap samples to consider for the estimation of the standard errors, $R$. Practitioners are always recommended to set \texttt{R} to the value that ensures the stability of results, bearing in mind that the higher the model complexity the larger $R$ is expected to be.}

{Further updates of the package will include the possibility to deal with multivariate responses in the spirit of \cite{Alfo2021}. Here, outcome-specific random coefficients are considered to model the unobserved heterogeneity characterizing each response variable. The corresponding multivariate distribution is left unspecified and estimated directly from the data. This proposal can be extended to the TV setting, as well as, to the mixed one (based on both TC and TV random coefficients) and implemented in the package. The inclusion of observed covariates on the latent layer of the model (as in mixtures of experts models) represents a further extension we aim at working on. }

\section*{Acknowledgments}
The work of Ranalli has been developed under the support of ``Fondo Ricerca di Ateneo, Universit\`a degli Studi di Perugia, edition 2021, project: AIDMIX''. The work of Marino has been supported by the ``Department of Excellence'' projects, 2018-2025 and 2023-2027, from Italian Ministry of Education, University and Research. The work of Salvati has been supported by the grant ``PRIN 2025 PNRR Prot. P2025TB5JF: Quantification in the Context of Dataset Shift (QuaDaSh)'' from Italian Ministry of Education, University and Research. {"The work of  Alf\`o has been supported by Sapienza university grant n. RG124191029C1395 ``Latent variable models for complex health data''.}

\end{document}